\documentclass[prb,aps,showpacs,prb,amsmath,twocolumn,floatfix]{revtex4}
\usepackage{graphicx}
\newcommand{\eq}[1]{Eq.~(\ref{#1})}
\newcommand{\fig}[1]{Fig.~\ref{#1}}

\def\figsize{0.48}

\begin{document}
\title{Transient currents and universal timescales 
for a fully time-dependent quantum
dot in the Kondo regime}
\author{Martin Plihal and David C. Langreth}\affiliation{Center 
for Materials Theory, 
Department of Physics and Astronomy, 
Rutgers University, Piscataway, NJ 08854-8019}
\author{Peter Nordlander}\affiliation{Department of Physics 
and Rice Quantum Institute,
Rice University, Houston, Texas 77251-1892 }
\date{21 September 2004}

\begin{abstract}
Using the time-dependent non-crossing approximation, 
we calculate the transient response of the current through a
quantum dot subject to a finite bias when the dot level is
moved suddenly into a regime where the Kondo effect is present.
After an initial small but rapid response, the time-dependent
conductance is a universal function of the temperature,
bias, and inverse time, all expressed in units of the Kondo
temperature.  Two timescales emerge: the first is the time
to reach a quasi-metastable point where the Kondo resonance
is formed as a broad structure of half-width of the
order of the bias; the second is the longer time required
for the narrower split peak structure to emerge from the previous
structure and to become fully formed.
The first time can be measured by
the gross rise time of the conductance, which does not substantially
change later while the split peaks are forming.  The second time
characterizes the decay rate of the small
split Kondo peak (SKP) oscillations in the conductance,
which may provide a method of experimental access to it.
This latter timescale is accessible via linear response
from the steady state
and appears to be related to the scale identified
in that manner
[A. Rosch, J. Kroha, and P. W\"olfle, Phys.\ Rev.\ Lett.\
\textbf{87}, 156802 (2001)].
\end{abstract}


\maketitle

\section{Introduction}
The theoretical predictions 
\cite{NgLee88PRL,glazman88, HershfieldetAl91PRL} of
consequences of the Kondo effect
for the steady state conduction through 
quantum dots began a decade ago.
At low temperatures, a narrow resonance in the dot density
of states can form at the Fermi level, leading to a large
enhancement of the dot's conductance, which is strongly dependent
on temperature, bias, and magnetic field.
Many of these effects have been recently observed by a
set of beautiful experiments by several groups.
\cite{GoldhaberetAl98Nature,GoldhaberetAl98PRL,CronenwettetAl98Science} 
These successes, supplemented by the
anticipation  that time dependent experiments are
not far behind, have spurred a number of theoretical groups.
\cite{HettlerSchoeller95PRL,SchillerHershfield96PRL,Ng96PRL,NordlanderetAl99PRL,GoldinAvishai98PRL,KaminskyetAl99PRL}
to consider the effects expected when
sinusoidal biases or gate potentials are applied. Indeed
recent experiments\cite{kogan}  have now seen Kondo sidebands.
Also the predictions \cite{aguado} of split Kondo conductance
peaks have been
observed in double \cite{albertchang} and multiple \cite{marcus} 
dots.
Surprisingly, the application of steps or pulses, which can provide
a less ambiguous measure of time scales than ac modulation,
have  been considered
less extensively theoretically,
\cite{NordlanderetAl99PRL,PlihaletAl00PRB,SchillerHershfield00PRB}
and not at all to our knowledge experimentally.
When pulsed voltage is applied to the to the quantum dot level
so that it suddenly is shifted into the Kondo regime, the conductance
of the dot will begin to increase. The current saturates when the system
reaches its new equilibrium configuration. In a previous 
investigation,\cite{NordlanderetAl99PRL}
we considered  a quantum dot biased by a small voltage and calculated
the time dependent change in linear response conductance when
a  stepped potential was applied to a gate, thereby shifting the
dot into the Kondo regime. Some general qualitative observations
were made, which now can be made quantitative through the
study of a different configuration of voltage switching.

In the present work we consider the
response of a quantum dot, operating as a single-electron transistor, 
already subject to a finite dc bias
when the dot level is shifted into the Kondo regime.
When a finite bias is present across the  dot, Kondo resonances appear at
each of the leads. 
Due to the finite bias, these resonances are broadened
compared to the zero bias situation.\cite{WingreenMeir94PRB}
While the equilibrium or zero bias situation is relatively well
understood, \cite{Bickers87RMP} much less is know about
the fully time-dependent situation, with Refs.~\onlinecite{rosch}
and \onlinecite{plihalweb} being most relevant to the present
work.

The present calculation determines within the non-crossing
approximation (NCA), the 
transient current after a gate pulse moves
the level of the single electron transistor into
the Kondo regime, under a large variety of temperatures
and biases. We are able to extract two different
timescales from our analysis, in addition to the trivial
very rapid scale of the very small time response
set by the tunneling rate to the leads (or the width
of the virtual-level resonance). The first of these
is characterized by the time to reach a quasi-metastable
point where the time-dependent conductance has essentially risen
to its equilibrium value, and where the Kondo resonance
has formed into a broad quasi-smooth structure of 
half-width roughly equal to the bias. The second is the timescale
for the emergence and formation of the
narrower
individual \textit{split} Kondo peaks (SKP),
at a much slower timescale.
This scale is also relevant for the decay of SKP conductance
oscillations, which may provide an experimental access to it.

\section{Model}
We model the quantum dot by a single spin degenerate level 
of energy $\epsilon_{\rm dot}$
coupled to leads through tunnel barriers.
The Coulomb charging
energy $U$ prevents the level from being doubly occupied. 

The system may be described by the following Anderson
hamiltonian:
\begin{equation}
H(t)=
\sum_\sigma \!
\epsilon_{\rm dot}(t)
n_\sigma +\sum_{k\sigma}\!\left[\epsilon_k%
n_{k\sigma}
+(V_k c^\dagger_{k\sigma}c_\sigma + {\rm H.c.})\right],
\label{hamiltonian}
\end{equation}
with the constraint that the occupation of the dot cannot
exceed one electron.
Here $c^\dagger_\sigma$ creates an electron of spin $\sigma$
in the quantum dot, with $n_\sigma$ the corresponding
number operator; $c^\dagger_{k\sigma}$ creates an electron in the leads.

For zero bias across the dot,
the general features of the static equilibrium spectral density when
the dot level $\epsilon$ is sufficiently below the Fermi level 
are well known. There is a broad resonance of 
half-width $\Gamma
  (\epsilon) = 2
  \pi \sum_k |V_k|^2 \delta(\epsilon -\epsilon_k)$
at an energy $\sim$ $\epsilon_{\rm dot}$.
The notation $\Gamma$ with no energy specified will always refer the
value at the Fermi level.  In addition, there is a sharp
temperature sensitive resonance at the Fermi level (the Kondo peak),
characterized by the low energy scale \cite{hewson}
 $T_K$ (the Kondo temperature),
\begin{equation}
T_K = D \left(\frac{\Gamma}{4\tilde E}\right)^\frac{1}{2}
 \exp\left(-\frac{\pi|\epsilon_{\rm dot}|}{\Gamma}\right),
 \label{tkondo}
\end{equation}
where $D$ is a high energy cutoff equal to half bandwidth when modeled by a 
symmetric flat band. Our calculations here use a symmetric parabolic
band of half bandwidth $D_0=9\Gamma$. We use
$D\simeq D_0/\sqrt{e} $, the choice that gives the
correct normalization for the leading  logarithmic 
corrections in the Kondo model. \cite{ShaoetAl194PRB} For our case where
$\epsilon_\text{dot}$ is in the band, $\tilde E=D$
in \eq{tkondo}. Only for $\epsilon_\text{dot}$ sufficiently below the
band cutoff does the form $\tilde E\propto |\epsilon_\text{dot}|$,
expected from the Schrieffer-Wolff transformation from the
Kondo model, 
result.
For finite bias $V$, the Kondo peak splits into two
sub-peaks \cite{ WingreenMeir94PRB} at $\pm V/2$ relative to 
the Fermi level, which we will always take to be at zero
energy. The nature of this splitting at large $V$
has recently been elucidated. \cite{rosch}

Our calculations  use  the non-crossing approximation (NCA),
which is reliable for temperatures down to
$T < T_\text{K} $ \cite{Bickers87RMP}. The details of the
time-dependent method of solution have been described in several previous
publications.\cite{LangrethNordlander91PRB,ShaoetAl194PRB}
Throughout this work energies, temperatures, and biases
are given
in units of $\Gamma$, and times in units
of $1/\Gamma$, with $\hbar=k_{\rm B} = e=1$.
In the regions of parameter space where $T_\text{K}$,
$T$, and $V$ are much smaller than $\epsilon_\text{dot}$
and $\Gamma$, physical properties are
functions of $T/T_\text{K}$, $V/T_\text{K}$ and $T_\text{K}t$, 
 alone, provided that the quantity under consideration
  is made dimensionless by appropriate factors of $T_\text{K}$
  and $G_0\equiv 2(e^2/2\pi \hbar)$.
In what
follows we mostly focus on properties relevant to this
universal region, 
except when a property is directly relevant for experiments.

In the present calculations,
we  investigate the transient electric currents trough the dot.
The  current into the dot  depends on the time $t$ as
\begin{equation}
I_{\rm in}(t)= 
ie\sum_{k\sigma}V_k \langle c_{k\sigma}^\dagger(t)c_\sigma(t)\rangle
+\mbox{c.c.}
\label{current}
\end{equation}
It may be divided into contributions $I_{\rm left}(t)$ and
$I_{\rm right}(t)$
by respectively restricting the $k$ summations to the appropriate lead.
For simplicity, we will only consider dots with
left-right symmetry.
The transport current is then 
 $I(t)=\frac{1}{2}[I_{\rm left}(t)-I_{\rm right}(t)]$.
The finite bias on the leads is taken into account by introducing a
time-dependent phase in $V_k$ in (\ref{hamiltonian}).
We calculate the Keldysh propagators corresponding to the
angular-bracketed expectation values in (\ref{current}) for each lead,
and hence obtain $I(t)$.

\section{Time-dependent conductance results}
	In this work our studies all start from an equilibrium
steady state 
with a bias across the dot equal to $V$,
but where the virtual level parameter $\epsilon_\text{dot}(t)$
is at a  negative value of magnitude sufficient that the
initial Kondo temperature is much smaller than the physical
temperature $T$, and the initial conductance is so small
as to be negligible.  In practice we use a starting value for
time $t<0$ of $-5\Gamma$.  At $t=0$ this virtual level
parameter is suddenly shifted to its final value,
which we will simply call $\epsilon_\text{dot}$ (with no
time argument), with the bias $V$ and virtual-level
width parameter $\Gamma$ unchanged.
The majority of our calculations take $\epsilon_\text{dot}$
to be $-2 \Gamma$, and we will denote the system
so described as \textit{system one}
(S1).
 Its Kondo temperature
$T_\text{K}=0.0022\Gamma$.  We also make a number
of calculations where $\epsilon_\text{dot}=-2.225\Gamma$,
for which $T_\text{K}=0.0011\Gamma$. This latter
system we call \textit{system two}
(S2).

When the dot level is in its lower
position, the Kondo temperature of the dot
is much smaller than the system temperature
and the Kondo resonance is essentially absent. The spectral function is
dominated by the broad virtual level of width $\sim 2\Gamma$ centered 
roughly at $\epsilon_\text{dot}(t<0)$. 
When the level is moved, a new virtual-level resonance of width $\sim 2\Gamma$
is formed around the new dot level. The time scale for the formation of this
resonance is $\Gamma$.\cite{NordlanderetAl99PRL}
The the Kondo resonance take longer time to form. In the $G(t)$
curves shown in this section, only this Kondo time scale is
apparent.

\subsection{Results for small bias\label{smallbias}}
We display first the results appropriate to very small $V$,
i.e. $V \ll T_\text{K}$, such that the conductance is
in the linear response regime, for a variety of temperatures
in system one.  These are shown in Fig.~\ref{figures/Gtt}.
It is obvious by inspection of the figure that the rise time
of the conductance, however that should be
defined,
is increasing as the temperature is lowered.
This
increase saturates when the temperature gets to around
$T_\text{K}$ or lower. More quantitatively the inverse
rise time (rise rate) is around $T_\text{K}$ at low temperatures,
but increases with $T$ for higher ones. These rates are
much slower than the rate $\Gamma$ which is associated with
the virtual level.
\def\figpath{figures/Gtt}
\begin{figure}[htb]
\centerline{\includegraphics[width=\figsize\textwidth]{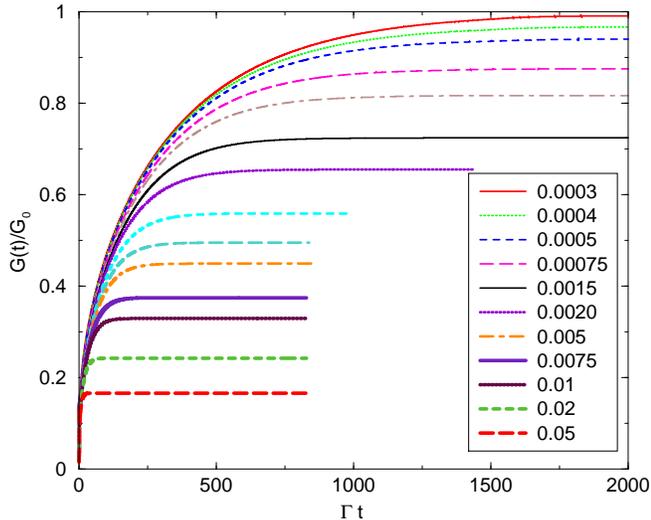}}
\caption{\label{\figpath}
Time dependent conductance vs.\ time for system one. The ordinate is
the time dependent conductance $G(t)\equiv I(t)/V$
in units of the open-channel conductance $G_0\equiv 2(e^2/2\pi\hbar)$.
As before, $I(t)$ is the time dependent current, and $V$ the bias.
The abscissa is the time $t$ after the gate switches
the dot into the Kondo regime, in units of $\Gamma^{-1}$.  The
numbers in the legend are the temperatures $T$ for the 
corrsponding curves in units of $\Gamma$.
 All curves are in the linear response
region of approaching zero bias $V$.
}
\end{figure}

In Fig.~\ref{figures/ucheck} we combine results from
system one and system two to test universality.
We find the results satisfying.  However, we believe that
the small differences between the two sets of curves are
not due to computational inaccuracy, but rather are
true deviations from universality,
however tiny.  The small time region $\Gamma t \sim 1$
is always non-universal, and
around 25\% smaller
for system two
than for system one.
With our present algorithms, we cannot further separate the
time scales
using
much smaller Kondo temperatures, 
so the results even on
the Kondo time scale are still slightly affected
by the nonuniversal fast contribution.
For this reason, the conductances for system two will be a little
smaller than for system one.
 In Fourier
space, one could say that there is a small non-universal
background, whose changes from system one to two are reflected
in the final value of $G$.  However, our results clearly
show that the parts expected to be universal
behave
in this manner.
\def\figpath{figures/ucheck}
\begin{figure}[htb]
\centerline{\includegraphics[width=\figsize\textwidth]{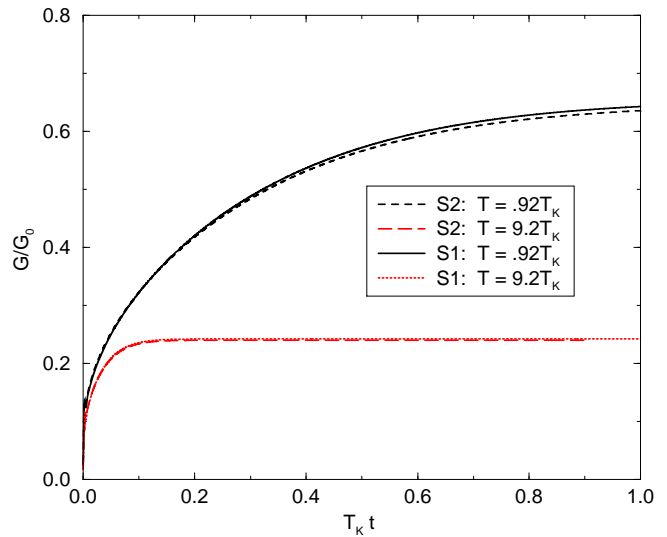}}
\caption{\label{\figpath}
Universality of time-dependent conductance curves. The S1 curves are
identical to the corresponding ones in Fig.~\ref{figures/conductanceV=0}
with the time axis rescaled to inverse $T_\text{K}$ units,
where here $T_\text{K}=0.0022\Gamma$. The S2 are curves for system
two (see text), which has a smaller $T_\text{K}$.
}
\end{figure}

\subsection{Results for finite values of bias}
The finite bias calculations were performed in the same way,
except 
that a constant
bias $V$
is present at all times.
In Fig.~\ref{figures/Gtv} we show the results for
system one for a variety of different biases and two
different
temperatures.
\def\figpath{figures/Gtv}
\begin{figure}[htb]
\centerline{\includegraphics[width=\figsize\textwidth]{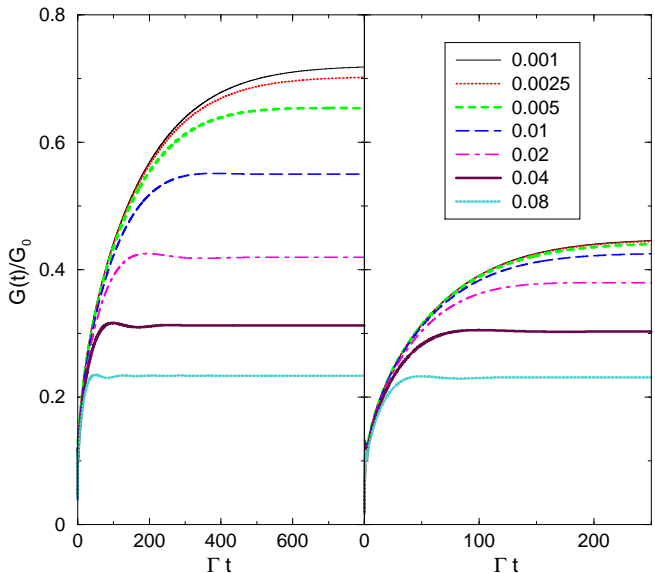}}
\caption{\label{\figpath}
Time dependent conductance vs.\ time. The ordinates are
the time dependent conductance $G(t)\equiv I(t)/V$
in units of the open-channel conductance $G_0\equiv2(e^2/2\pi\hbar)$.
Here $I(t)$ is the time dependent current, and $V$ the bias.
The abscissas are the time $t$ after the gate switches
the dot into the Kondo regime, in units of $\Gamma^{-1}$. Both
panels are for system one (see text), with $T=0.0015 \Gamma$
(left panel) and $T=0.005 \Gamma$ (right panel). The
numbers in the legend are the biases $V$ 
in units of $\Gamma$ for the 
corrsponding curves in each panel. 
}
\end{figure}
The simple descriptive facts that are evident by inspection of
the rise-times implied by these
curves
are
that: (i) for $V<\sim T$ there is little change;
the rising rate
proportional to $T$ discussed in the previous
subsection \ref{smallbias} still prevails;
(ii) for  $V>\sim T$ the rising rate increases roughly linearly
with $V$.

On a finer scale, one may see small oscillations about the
final value, especially in the curves for larger $V$.
These result from the fact \cite{WingreenMeir94PRB,Sivan}
that in the presence of $V$ the Kondo peak in steady
state is split into two peaks at $\pm V/2$ respectively.
These split Kondo peak (SKP) oscillations were clearly
identified in  earlier work.
\cite{PlihaletAl00PRB,SchillerHershfield00PRB,ColemanUnstable}
They have a frequency almost precisely equal to $V$
and a lifetime that is much longer than that implied
by the $V$ and $T$ rate scales, something suggested by
the prescient fact that in mean field theory the oscillations
are undamped. \cite{ColemanUnstable}  We put off a detailed
analysis to a later section.

\subsection{Initial oscillations}
For small times, $\Gamma t < 10$ in Figs.~\ref{figures/Gtt} 
and \ref{figures/Gtv}
there appear oscillations, which are invisible 
on
the scale
of those figures. As opposed to the features mentioned
in the previous two subsections,
they are non-universal in the usual sense.  However,
 for a given system, they take the same form and
frequency, independently of $T$ and $V$, at least for
values of these parameters that are much smaller than
$\epsilon_\text{dot}$ or $\Gamma$.  They are shown in
the top curve in the top panel of Fig.~\ref{figures/initialwig}
\def\figpath{figures/initialwig}
\begin{figure}[htb]
\centerline{\includegraphics[width=\figsize\textwidth]{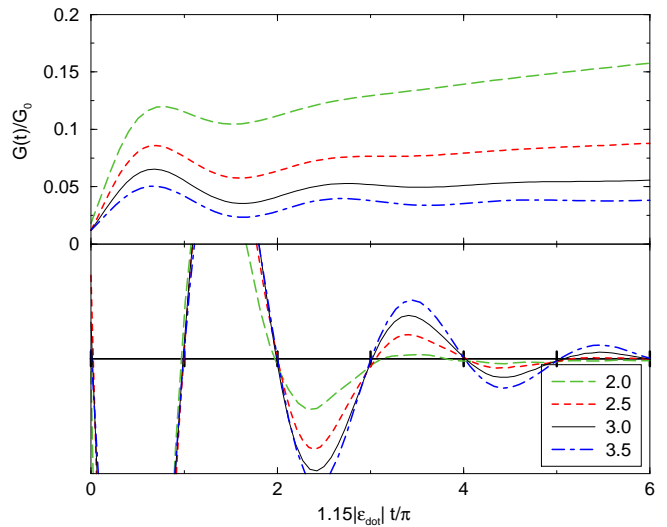}}
\caption{\label{\figpath}
Initial oscillations at short times. The top (long-dashed) curve
in the top panel is the short time version of the system one
($|\epsilon_{\rm dot}|=2\Gamma$) curve for bias $V=0.01\Gamma$. Thus
it represents the very short time behavior of the curves in
Fig.~\ref{figures/Gtv}.
The behavior is similar for all biases studied.
The other
three curves are for other values of  $|\epsilon_{\rm dot}|$
whose ratio to $\Gamma$ is indicated in the legend. In each
case the time coordinate is rescaled as indicated, to show
that the frequency of the oscillation is about equal to
the magnitude of virtual resonant level's energy (measured from
the Fermi level).
The renormalized value of
the latter is a bit larger than the bare value $|\epsilon_{\rm dot}|$.
The bottom panel, showing the second derivatives of the curves in
the top panel, more clearly identifies the period of oscillation.
}
\end{figure}
for system one at $T=0.0015 \Gamma$ and $V=0.01\Gamma$.
The next several curves show the initial oscillations
for successively larger values of $|\epsilon_\text{dot}|$.
The second panel more clearly establishes that the
frequency of these oscillations is the difference between
the Fermi level and the virtual level position.

\subsection{Alternative roughly equivalent measurement}
The investigations described above shows that by measuring the transient currents
in a quantum dot subject to a finite bias as a function of bias
$V$
and
temperature, one can probe the bias-induced and thermal broadening 
of the  non equilibrium Kondo problem. The most straightforward implementation
of such a measurement would be to measure the total charge transported through
the dot $Q(\tau)$ when the dot level is subject to a pulse train in which
the dot level is raised to the Kondo regime for a time $\tau$.
By measuring the total charge $Q(\tau)$ transported trough the dot
during a single pulse as a function
of pulse length $\tau$ and taking the derivative with respect to $\tau$,
$\frac{dQ(\tau)}{d\tau}$,  a quantity is obtained that closely
follows the transient currents in the dot.  The equivalence would
be exact if the the current were instantly returned to zero
at the end of the pulse. Since the switching goes to a non-Kondo
region, the turnoff rate will be $\sim\Gamma$. Features that
occur on that scale like the initial oscillations will be
masked, but the $V$ and $T$ rates, as well as the SKP oscillations
will be preserved.  We have verified this by explicit calculations.

\section{Analysis and interpretation of zero bias results}
\subsection{Large time limit}
In Fig.~\ref{figures/gequil} we display the large time
limit of our calculations for systems one and two. Since
the bias is essentially zero, these
represent the
steady state linear response conductance. 
This figure shows that our results agree with general expectations
including approximate universality.

The exact asymptotic
curve  at large $\ln (T/T_\text{K})$
\begin{equation}
\frac{G}{G_0}=\frac{3\pi^2}{16\ln^2 ({T}/{T_\textrm{K}})}
\label{eq:abrikosov}
\end{equation}
was first calculated by
Abrikosov \cite{abrikosov} for the Kondo impurity problem,
and has been more recently adapted and applied
applied to quantum dots.
\cite{acdriving}  It can also be derived by the
so-called poor man's scaling method. \cite{poorman,kaminskiPRB}
 However, if one applies
the perturbative procedure used in Ref.~\onlinecite{rosch}
for large $V$ to the similar large $T$ case, one finds
that within NCA, the asymptote is $4/3$ times the value
in \eq{eq:abrikosov}.  The NCA asymptotic is the one shown
in the figure.
Our lowest temperature point
(not shown)
is slightly above the unitarity limit, another  NCA error also
found by others. \cite{rosch}
\def\figpath{figures/gequil}
\begin{figure}[htb]
\centerline{\includegraphics[width=\figsize\textwidth]{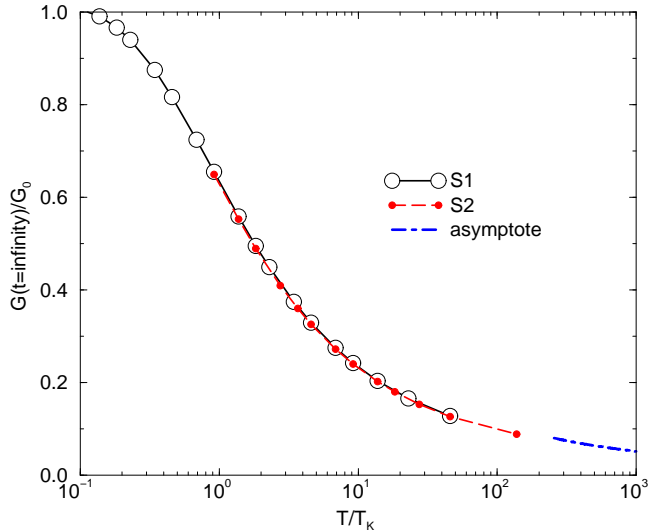}}
\caption{\label{\figpath}
The large-time limit of $G(t)/G_0$ vs. temperature. Since
$V\sim 0$, this is the equilibrium value of $G/G_0$.
The points
labeled S1 and S2 are from system one and two, respectively
(see text). The curve labeled  \textit{asymptote} in the
legend represents the large $T$ asymptote within NCA
(see text).
}
\end{figure}

\subsection{Extraction of the rise rate}
Here 
we
 use previous analytic results
\cite{NordlanderetAl99PRL} to extract the rise
time $\tau$ from the data of 
Fig.~\ref{figures/Gtt}
What was found there, was that the current response to a stepped
turning on of the Kondo interaction
would be the same as  the  {\it equilibrium}
response to a time-dependent {\it effective temperature } $T_{\rm e}$
given by $ T_{\rm e}=T \coth \pi T t/2\hbar$. We start with the
tautology 
$
\delta I/I = [d(\ln G)/d(\ln T_{\rm e})]\; \delta T_{\rm e}/ T_{\rm e}.
$
Defining the fraction $f$ as the finite difference 
\begin{equation}
f \equiv  \frac{I(\infty)-I(t)}{I(\infty)},
\label{f}
\end{equation}
and
 \begin{equation}
f_0\equiv-2\,\frac{d(\ln G)}{d(\ln T)},
\label{f0}
\end{equation}
we have  in the large $t$ limit
$
 f  \rightarrow  f_0(\coth \pi Tt/2\hbar - 1 )/2.
$
which becomes
\begin{equation}
 f \rightarrow  
	f_0  \exp \left(-\frac{t}{\tau}\right),
\label{dell}
\end{equation}
with
\begin{equation}
\frac{1}{\tau}=\pi T
\label{alpharate}
\end{equation}.

This suggests that we should fit the upper part of the our
curves in Fig.~\ref{figures/Gtt} to \eq{dell}, and compare
the resulting $\tau$ with \eq{alpharate}. Of course,
the derivation in Ref.~\onlinecite{NordlanderetAl99PRL}
was only to lowest logarithmic order, and certainly
cannot be expected to be valid for $T <\sim T_\text{K} $,
and we might also expect the possibility of  logarithms of 
$T/T_\text{K}$ to become predominant at large $T$.
Nevertheless, we find no evidence for the latter
effect
in the curves of Fig.~\ref{figures/Gtt}. We display the rates
($\equiv 1/\tau$) that we find in Fig.~\ref{figures/conductanceV=0}.
\def\figpath{figures/conductanceV=0}
\begin{figure}[htb]
\centerline{\includegraphics[width=\figsize\textwidth]{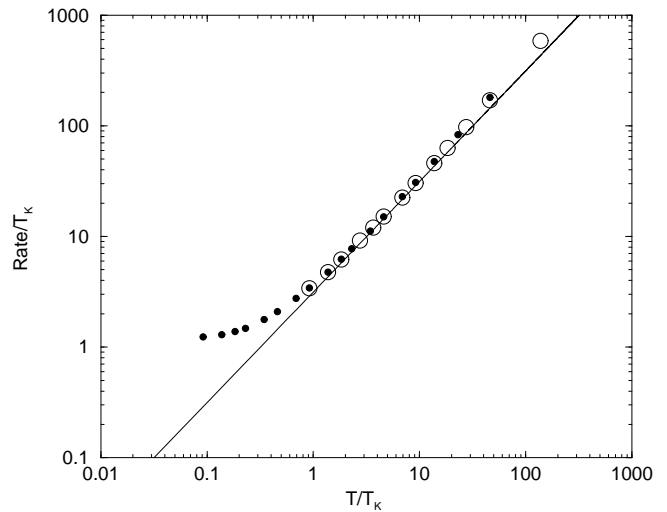}}
\caption{\label{\figpath}
Rate ($1/\tau$) at which the conductance approaches
its final value versus temperature $T$, with zero bias $V=0$.
The solid 
dots are for system one (see text), while the open circles
are for system two, which has a different Kondo temperature
$T_{\rm K}$. The solid curve is a straight line of slope
$\pi$ through the origin (see \eq{alpharate}). 
}
\end{figure}
The quite good agreement of the curves in 
Fig.~\ref{figures/conductanceV=0} over a major part of the
range is satisfying.  The curves also strongly suggest
that the low temperature limit of  \eq{alpharate}
replaces $T$ by something of the order of $T_\text{K}$.
Our points appear to deviate from \eq{alpharate}
at higher temperatures.  However, our fitting procedure
is less accurate at high temperatures where $G(\infty)$
is small, 
and
our estimated
numerical
error is of the
same magnitude as this deviation.
For this reason,
we are
not prepared to say with certainty whether this is
a real effect or not.
An analysis of the temporal evolution of the instantaneous spectral function
of the dot level does reveal significant spectral reshaping of the Kondo resonance well
beyond the time where the conductance has saturated.\cite{NordlanderetAl99PRL}
This effect suggests the presence
of a slower time scale at large temperatures.
We might
certainly expect
a slower timescale to emerge,
if we apply
the arguments of Ref.~\onlinecite{rosch} to the low bias,
high temperature case, as done in Appendix~\ref{sec:KORRINGA}.
A more detailed investigation of the origin of
this effect and the possibility 
for its experimental detection is in progress.

\section{Analysis and interpretation of finite bias results}
\subsection{Large time limit}
We begin by plotting the large time values of $G(t)$
in our finite $V$ calculations. This of course is
the steady state conductance, and our curves in
Fig.~\ref{figures/Gsaturated} supplement results
\def\figpath{figures/Gsaturated}
\begin{figure}[htb]
\centerline{\includegraphics[width=\figsize\textwidth]{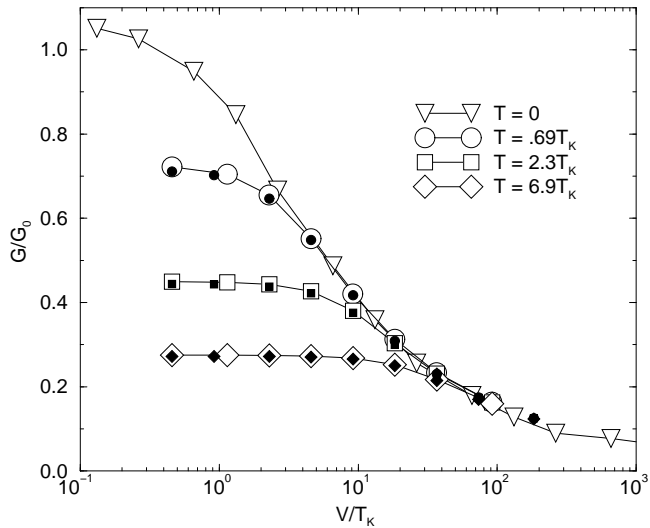}}
\caption{\label{\figpath}
Long time or dc conductance vs.\ bias.  The ordinate $G/G_0$
is the dc conductance in units of the open channel conductance
$G_0=2e^2/2\pi\hbar$ for the two spins, while the abscissa is
the bias $V$ in units of the Kondo temperature $T_{\rm K}$.
The triangles represent the data of Ref.~\onlinecite{rosch}
with the abcissa rescaled to correct for the $\sim 30\%$  
difference between the two definitions of the Kondo temperature.
The open circles, squares, and diamonds represent the
long time saturation of our time-dependent conductance curves
for system one at the temperatures indicated.  The smaller solid
symbols represent the values at the same respective temperatures
for system two, which as a different Kondo temperature (see text).
}
\end{figure}
previously calculated \cite{rosch} for $T=0$, which
are included in the plot. Since the formula for
the Kondo temperature used there gave a value
some 30\% greater than \eq{tkondo}, their  values
for $V/T_\text{K}$ were rescaled appropriately, so they
could be put on the same graph with ours.
Our values show approximate universality as discussed
earlier and agreement with Ref.~\onlinecite{rosch}
in the appropriate regions.

\subsection{Extraction of the rise rate}
The very large time behavior of our curves in
Fig.~\ref{figures/Gtv} is dominated by the
decay of the SKP oscillations.
It is clear that their
amplitude is very small and for the smaller $V$'s entirely
negligible, and they will have
no effect on any
common sense definition or experimental measurement
of rise time. Despite the SKP oscillations, the form
\eq{dell}
well describes the upper part of
the time-dependent conductance curves in \fig{figures/Gtv}
up until they reach within $\sim 1\%$ of the saturation value.
Therefore, as a
practical empirical technique to characterize
the rise time,
we use the same technique as for $V=0$.
The results  of this analysis are shown in \fig{figures/conductanceV}.
\def\figpath{figures/conductanceV}
\begin{figure}[htb]
\centerline{\includegraphics[width=\figsize\textwidth]{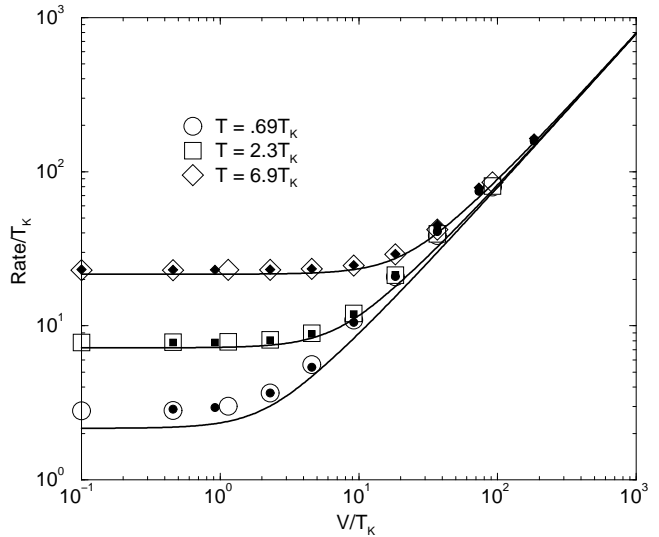}}
\caption{\label{\figpath}
Rate (inverse time) at which the conductance approaches
its 'final value' versus bias $V$, at several temperatures
as indicated. The open circles, squares, and diamonds
are for system one (see text), while the corresponding
solid symbols are for system 2, which has a different
Kondo temperature. Our $V=0$ points are
shown in the figure at $V/T_{\rm K}=0.1$. 
The solid lines are the predictions
of \eq{eq:korringa2}.
}
\end{figure}
\subsection{Interpretation\label{interpretaion}}
If the source of the $T$ in \eq{alpharate} is the same
as the source of $T$ in the Korringa rate, \eq{eq:korringa},
then it
 arises because a change
in $\vec S$ involves the  absorption of an
electron-hole pair of zero energy from the leads.  
The phase space for such
a process
is doubly restricted by the Pauli principle, and
produces the factor
$
\int \! d\epsilon f(\epsilon)(1-f(\epsilon))
= T
$
where $f(\epsilon)$ is the Fermi function.  

The application of these ideas to a biased  dot allows
the phase space factor to be tuned in a continuous fashion
by varying the bias $V$ across the two leads of the dot,
which for simplicity we assume to be symmetric under lead interchange.
The phase space restriction factor in this case is
given by 
$\frac{1}{4}\sum_{l,l'}\int\! d\epsilon\; f_l(\epsilon)(1-f_{l'}(\epsilon))$,
where the indices $l$ and $l'$ designate which lead is referred to.
For example, the Fermi functions $f_1(\epsilon)$  and $f_2(\epsilon)$
have Fermi levels displaced by 
 $\pm e{V}/{2}$, respectively,
where $e$ is the magnitude of the electronic charge and $V$ is the bias.
The integral above can be evaluated analytically, with the result
that  Eq.~(\ref{alpharate}) should be replaced by
\begin{equation}
\frac{1}{\tau}= \pi T F\left(\frac{V}{T}\right),
\label{eq:korringa2}
\end{equation}
where
\begin{equation}
F(x)= \frac{1}{4}\left(1+1 + \frac{x}{1-e^{-x}}
	+\frac{xe^{-x}}{1-e^{-x}} \right).
\label{eq:F}
\end{equation}
In writing Eq.~(\ref{eq:F}) we have sacrificed conciseness
to facilitate clarification of the origin of the terms in parentheses,
in terms of the wave function of the annihilated particle-hole pair.
The first two terms arise when the two components of this object
are in the same lead; in this case the existence of $V$ has no
effect on the result;
 the phase space for these processes
is still constricted,
and the contribution to $1/\tau$ is  still small.  For the third
term, the particle is on lead 1 and the hole on lead 2; here
the phase space is opened wide by $V$.  Finally for the
fourth term, the particle is on lead 2 and the hole on lead 1;
the phase space is, aside from an exponentially small tail,
closed off entirely as $V$ is increased.  So the essential
physical feature deriving from Eq.~(\ref{eq:F}) is that
the factor of $ T$ in Eq.~(\ref{alpharate}) is replaced at
large $V$ by $\frac{1}{4}V$. The notion of an expanded phase
space is implicit in the Anderson model rate calculation of
 Wingreen and Meir \cite{WingreenMeir94PRB} and in
a different context in the work of
 Kaminski {\it et al.} \cite{KaminskyetAl99PRL}
In any case, the rate of \eq{eq:korringa2} provides a way
to rationalize the calculated points in \fig{figures/conductanceV}
in a parameter free way, which has predictive power for
the rise-time of the conductance.
The comparison shown in that figure shows that it captures
the main trends of the NCA
results, although not with such
good agreement as for the $V=0$ case (\fig{figures/conductanceV=0}).
\section{Time-dependent spectral function---two timescales}
\subsection{Faster timescale}
As an aid to the interpretation of the conductance behavior,
we display the time-dependent spectral functions $A_\text{dot}(\omega,t)$
for the dot.\cite{acdriving}
\def\figpath{figures/spec1}
\begin{figure}[htb]
\centerline{\includegraphics[width=\figsize\textwidth]{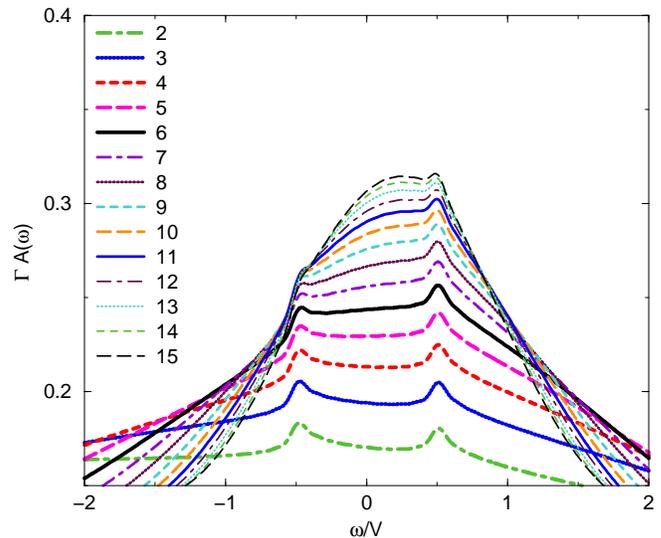}}
\caption{\label{\figpath}
Time-dependent spectral functions for the time scale relevant
for the rise time of the conductance. These curves are for
system one with $T=0.0015\Gamma = 0.69T_\text{K}$, and
$V=0.08\Gamma=37T_\text{K}$. The curves are at equal time intervals
$\Delta t=2.8\Gamma^{-1}=0.006T_\text{K}^{-1}=0.17(4/\pi V)$,
where the $4/\pi V$ factor is motivated by the high temperature asymptote
of \eq{eq:korringa2}.  This factor has the
 value of unity at $t=6\,\Delta t$
(thick solid black curve). The lowest curve is for time
($t=2\,\Delta t$) after the virtual level parameter was switched to
it's final value of $-2\Gamma$. By curve $15$ the conductance
has for all practical purposes reached its dc value.
}
\end{figure}
\fig{figures/spec1} shows the case $T=0.0015\Gamma=0.69T_\text{K}$
and $V=0.08\Gamma=37T_\text{K}$. The snapshots shown here
are for the time scale appropriate for the rise time of
of the conductance, as shown in detail in \fig{figures/exp}.
\def\figpath{figures/exp}
\begin{figure}[htb]
\centerline{\includegraphics[width=\figsize\textwidth]{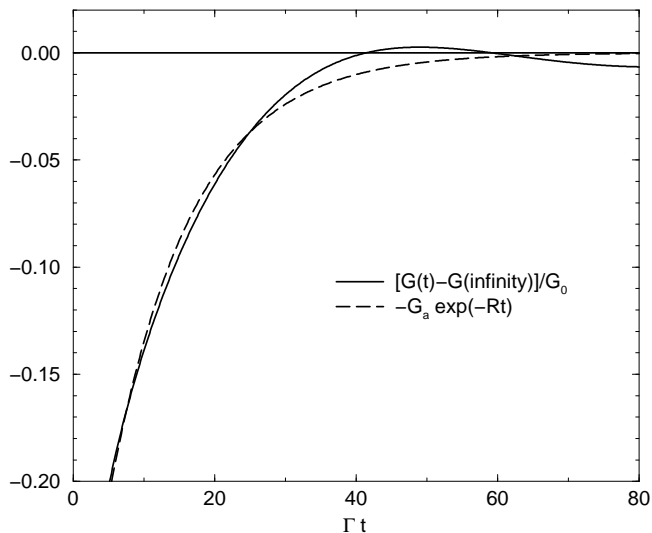}}
\caption{\label{\figpath}
The rise in conductance as the Kondo peak forms for bias
$V=0.08\Gamma$ at $T=0.0015$.  The solid curve is a magnified
section of the corresponding curve in Fig.~\ref{figures/Gtv}
with the horizontal axis shifted as indicated in the legend.
The dashed curve is the indicated exponential fit. The rate
$R$ obtained in this way characterizes the rise time
of the conductance, and this is the rate used in the preparation
of Fig.~\ref{figures/conductanceV}.  Of all the curves
in Fig.~\ref{figures/Gtv} this is the worst case for the
exponential fit, because here the SKP oscillations are largest.
The first period of these oscillations is partially visible
here.  Multiple cycles are shown in the larger magnification
of Fig.~\ref{figures/skposc}.
}
\end{figure}
\fig{figures/spec1} shows the rise of the Kondo peak as
a mostly smooth structure of half-width $\sim V$, with
the individual split peaks mostly undeveloped.  This structure
appears to be converging toward a quasi-stationary steady
state. It is clear that this timescale, intermediate between
the trivial very rapid timescale governed by the $\Gamma$ rate,
and the longest timescale yet to emerge, is the one governing
the rise time of the conductance. It is probably only available
through a fully non-equilibrium theory, since it is the
scale relevant for approaching a steady state that is 
metastable at best. It is therefore probably unavailable through
perturbation theory from the true steady state, and
hence will not likely appear in steady-state correlation functions.

\subsection{Slower timescale}
\fig{figures/spec2} shows the development of the spectral function
on the longest timescale, showing the rate at which the
individual split peaks emerge from the
quasi structure of width $V$
formed at the metastable point.  For $V \gg T_\text{K}$, this
latter rate is much smaller than that illustrated in 
\fig{figures/spec1}, in this case by an order of magnitude.
During the whole course shown in \fig{figures/spec2},
the conductance changes only by the small oscillatory amounts
indicated by the SKP oscillations. Indeed, the area under
$A_\text{dot}$ (when $T \ll V$) in
between $-V/2$ and $V/2$ is a virtual constant in time
in this time regime, with the additional area under the
split peaks being almost exactly compensated by the loss in
area between peaks, continuously in time.  Indeed,
we have verified that 
the conductance in this regime, which still has a tiny
fluctuation due to the SKP oscillations, is given by
the steady-state formula (for example, Eq.~(12) of
Ref.~\onlinecite{WingreenMeir94PRB}), provided that the $A_\text{dot}$
used is the time-dependent
instantaneous spectral function.\cite{nconst}. 
\def\figpath{figures/spec2}
\begin{figure}[htb]
\centerline{\includegraphics[width=\figsize\textwidth]{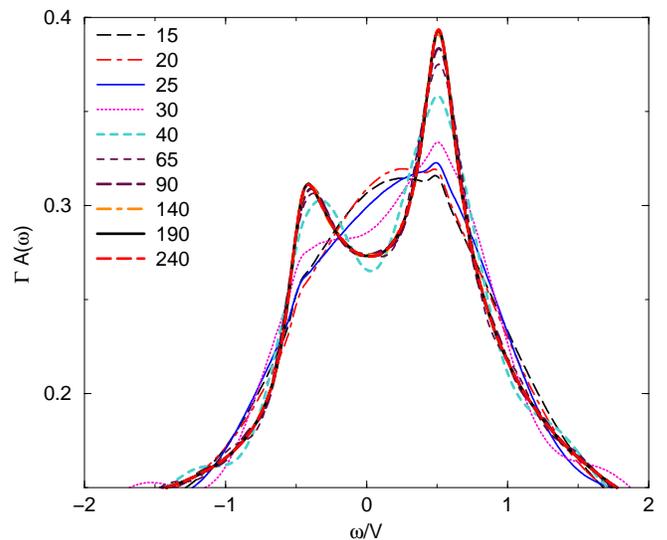}}
\caption{\label{\figpath}
Time-dependent spectral functions for the time scale relevant
for the development of the split Kondo peak and the
damping of SKP oscillations. This is a continuation
of the snapshots shown in \fig{figures/spec1}, so all
the parameters have the values shown in the caption of
that figure. Here, however, the time interval between
the successive snapshots is much greater, varying
from $5\,\Delta t$ to $50\, \Delta t$, as indicated
by the labeling in the legend.
}
\end{figure}
\section{SKP oscillations and the longest time-scale}
\subsection{SKP oscillations in the quantum dot}
In \fig{figures/exp}, which extends a little beyond the
range of \fig{figures/spec1}, one can see the beginnings of the
SKP oscillations.  The continuations of them on
a highly magnified scale are shown in
 \fig{figures/skposc}. 
\def\figpath{figures/skposc}
\begin{figure}[htb]
\centerline{\includegraphics[width=\figsize\textwidth]{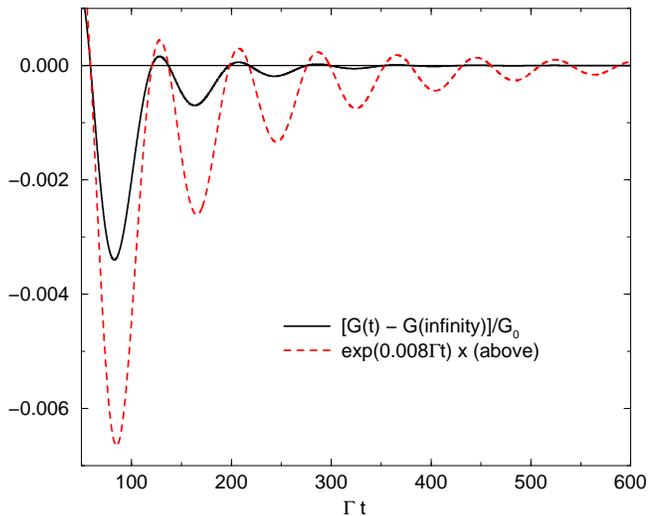}}
\caption{\label{\figpath}
SKP oscillations for bias $V=0.08\Gamma$ at $T=0.0015$.
The solid curve is a magnified section of the corresponding
curve in Fig.~\ref{figures/Gtv} with the position of
the horizontal axis shifted as indicated in the legend. The
dashed curve is a progressivly more magnified version of the solid
curve, so that more SKP oscillations can be seen.  As indicated
in Fig.~\ref{figures/positions}, the angular frequency of these
oscillations is almost exactly equal to $V$.
}
\end{figure}
Generally, many complete periods of oscillation can be seen,
and 
this period corresponds almost exactly to an
angular frequency of $V$, becoming closer and closer from
below as $V$ gets larger and larger. The details of this analysis
is given in Appendix \ref{SKPdetails}.  There it is also shown
that the decay rate is close to constant (exponential decay).
For large $V$ this rate is substantially smaller than
the rates identified earlier characterizing the rise time
of the conductance. The method of extraction of the rates
is described in Appendix \ref{SKPdetails}.

For the case shown here, this rate is on the order of the
distance to the inflection point on either of the peaks in
the converged spectral function.\cite{inflection}
The rate is
not inconsistent with,
but possibly a little slower than, the slow-scale rise rate
of the time dependent spectral function,
although
the latter
is
more difficult to pinpoint. We can with
much less ambiguity compare this rate with the $2\gamma$
rate identified in Ref.~\onlinecite{rosch}, and calculated
for $T=0$.  What we find is that if we divide the SKP oscillation
decay rates by two, then
these values at our lowest temperature ($\sim 0.3 T_\text{K}$)
agree within their accuracy with the $2\gamma$ curve. This is
shown in \fig{figures/skprates}. 
\def\figpath{figures/skprates}
\begin{figure}[htb]
\centerline{\includegraphics[width=\figsize\textwidth]{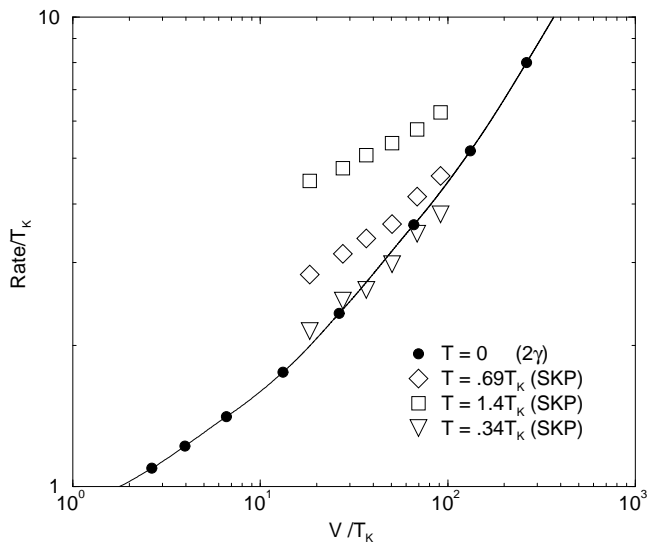}}
\caption{\label{\figpath}
Decay of SKP oscillations vs.\ bias $V$. The empty squares,
diamonds, and triangles
represent data derived from the damping rate of
the SKP oscillations in our calculations of conductance
vs.\ time  at the temperature indicated respectively.
The solid circles are 
the calculations of $2\gamma$ at zero temperature in
Ref.~\onlinecite{rosch}.
}
\end{figure}
We are not certain why 
half the decay rate is what seems to correspond to
$2\gamma$,
but presumably the issue is dephasing which occurs
both at $-V/2$ and $+V/2$ giving roughly additive contributions.
Another possibility is that $T=0$ (which is currently
unavailable for us) will be further from our lowest points
at $T=0.34 T_\text{K}$ than we expect.
But a definitive answer will have to await further and
more complete studies. What is very clear, however, is that
the damping of SKP oscillations is controlled by a rate
much slower than $\sim V$ for large $V$, and which is
numerically quite close to the $2\gamma$ rate.

\subsection{Decay of SKP oscillations in $N$-fold degenerate models}
We conclude this section with a connection with
models which increase $N$ from the $N=2$ appropriate for
quantum dots. Repeating the large $V$ NCA perturbative
analysis \cite{rosch} for that case gives
\begin{equation}
2\gamma = \frac{\pi V}{2N\ln^2 \frac{V}{T_\text{K}} 	}.
\label{ndecay}
\end{equation}
As noted in Appendix \ref{sec:KORRINGA}, the NCA \textit{does not}
need correcting by multiplying this by $S(S+1)$, which would
be proportional to $N^2$ at large $N$. \eq{ndecay}
indicates that the decay rate of SKP oscillations approaches
zero as $1/N$ for large $N$. This is consistent with the
trend of NCA studies \cite{PlihalLangreth}
 of this damping as a function of $N$, and with the
 fact \cite{ColemanUnstable} that these oscillations are
 undamped in mean field theory ($N=\infty$). In this case
 the effect of the vertex correction (see Appendix \ref{sec:KORRINGA})
 is dramatic indeed, making a qualitative change, as opposed
 to the $\sim 30\%$ correction for the $N=2$ quantum dot case.
 
\section{Additional conclusions}
Many of our important conclusions are contained in the abstract and
final paragraph of the introduction.
  We mention here the few
that are not.

First, for $V\sim 0$ the rise-time of the conductance is
very accurately characterized by the simple expression $\pi T$,
\eq{alpharate}, over a wide range of temperature $T>T_\text{K}$,
and appears to heal towards $T_\text{K}$ for smaller $T$. 
Second, for 
$V>~ ~4T$, one should replace $T$ by $V/4$ in the above.
These rates appear to lead toward a quasi-metastable point,
and do not negate the emergence of slower rates at longer
times, and which in the case of larger $V$ control the
damping of SKP oscillations. 
Third, at the very beginning
of switching the gate, there are small initial nonuniversal
oscillations at a frequency corresponding to the dot level's
separation from the Fermi level (for $V$'s small with
respect to that separation). Finally we predict for large
$V\gg T_\text{K}$ for $N$-fold degenerate models that
the damping of SKP oscillations decreases as $1/N$.
\acknowledgments
We thank A. Rosch for relevant communications.
The work was supported in part by NSF grants DMR 97-08499 
and DMR 00-93079 (Rutgers),
DOE grant DE-FG02-99ER45970 (Rutgers), 
and  
the Robert A. Welch foundation under grant C-1222 (Rice)

\appendix

\section{The Korringa rate\label{sec:KORRINGA}}
A half a century ago, the framework and starting point for
discussion of the time scale $\tau$ for a localized spin in
an electron sea was set by Korringa, \cite{korringa}
whose contribution has been promulgated at the textbook
level for decades. \cite{slichter}
  In simple terms  $1/\tau$ gives the
fractional rate at which a component of 
spin ${\bf S}$ representing a magnetic impurity 
(or a quantum dot) is  changing due to the
electrons in the conduction band (or in the leads of a quantum dot).
It is given by an expression of the type
\begin{equation}
\frac{1}{\tau_\text{kor}}= \alpha T,
\label{eq:korringa}
\end{equation}
where $ \alpha $  is a dimensionless constant.
and $T$ is the temperature. 
Although Korringa's
original derivation applied to a nuclear spin where the interaction
with electrons is dipolar, it has been widely applied to impurity
electron spins as well, using the Kondo model 
\begin{equation}
H=\sum_i( \epsilon_{k_i} - {J}{\bf{s}}_i\cdot{\bf S}),
\label{hkondo}
\end{equation}
for the interaction. 
In \eq{hkondo},
 $k_i$ is a generalized  quantum number for the 
$i^{\rm th}$ conduction (or lead) electron, $ \epsilon_{k_i}$ is its
energy, and $ {\bf{s}}_i$ is its spin operator.  For a 
symmetric quantum dot,
$k$ is assumed to include the information on which
lead is referred to.  The exchange coupling $J$ is taken to
be independent of $k$, aside from the usual high-energy cutoff
at energies  further than $D$ above or below the Fermi level.
In terms of Anderson model parameters $J=-2|V|^2/\epsilon_\text{dot}$,
where here $V$ is the matrix element in \eq{hamiltonian}, and not
the bias voltage.
To lowest order in $J$, the quantity $\alpha$ in Eq.~(\ref{eq:korringa})
is given, for example,  by  Eq.~(1.16) in 
Ref.~\onlinecite{LangrethWilkins},
\begin{equation}
\alpha=\pi (J\rho)^2,
\label{alphakor}
\end{equation}
where $\rho$ is the density of states in the leads (both
leads together), when applied to a symmetric quantum dot.

One should note that (\ref{alphakor}) applies to a general
spin ($S$) Kondo model, ($S=\frac{1}{2}$ for a quantum dot),
and \textit{there is no factor of} $S(S+1)$. For $S=\frac{1}{2}$
this can be shown to be incontrovertible  by the solution
of a simple master equation for the occupation probability
of up or down spin states
$\dot P_\sigma=-R(P_\sigma-P_{-\sigma})$, where 
$R=R_{\uparrow\downarrow}=R_{\downarrow\uparrow}$ is the 
spin-flip rate of the dot spin due to it's interaction
with the lead electrons.  This is readily solved for
the decay rate of an averaged spin component
$\langle\dot S_m\rangle = -\langle\dot S_m\rangle/\tau_\text{kor}$
where $1/\tau_\text{kor}=2R=\alpha T$, where $R_{\downarrow\uparrow}$
was evaluated by the Fermi golden rule, giving (\ref{eq:korringa})
with $\alpha$ given by (\ref{alphakor}).

While it is true that the equilibrium pseudo-fermion propagator's
self-energy's imaginary part \textit{does} indeed contain the
$S(S+1)$ factor [Eq.~(C4) in Ref.~\onlinecite{LangrethWilkins}],
the time decay rate of the spin correlator does not,
because of vertex corrections, which are summed in
Ref.~\onlinecite{LangrethWilkins} via the Kadanoff-Baym
equations. 

When $J$ is rescaled via the poor man's
scaling
technique,
$J\rho \rightarrow -1/\ln(T/T_\text{K})$, as done by
many authors to obtain results valid for
$\ln(T/T_\text{K}) \gg 1$, one gets
\begin{equation}
\frac{1}{\tau}=\frac{\pi T}{\ln^2 \frac{T}{T_\text{K}}}
\label{tauasymp}
\end{equation}
Similarly one gets large $V$ results through the replacement
$T\rightarrow\frac{1}{4}V$ in \eq{tauasymp}. These agree
exactly with what one gets from high $V$ or high
$T$ expansions\cite{rosch}  of the NCA equations,
and we conclude that NCA gets the right answer here.
Of course, the NCA misses the factor $\frac{3}{4}$ in the
conductance, \eq{eq:abrikosov}, as previously pointed out.\cite{rosch}

\section{Details of SKP oscillation parameters}
\label{SKPdetails}
In \fig{figures/positions} we map the successive positions of the
zero crossings as well as the positions of the 
\def\figpath{figures/positions}
\begin{figure}[htb]
\centerline{\includegraphics[width=\figsize\textwidth]{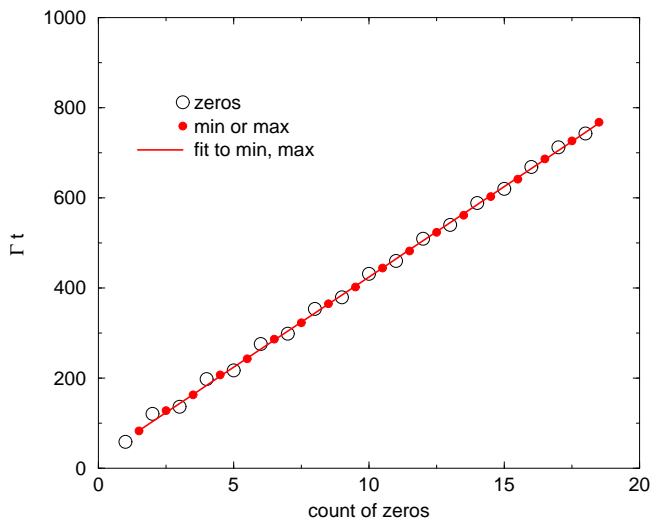}}
\caption{\label{\figpath}
Positions of extrema and zeros of SKP oscillations for $V=0.080\Gamma$
as in Fig.~\ref{figures/skposc}. The slope of the fitted
line gives a frequency of $0.0783\Gamma$. The frequency is always
a speck less than $V$, but for our largest $V$'s, we find
the difference to be only around $0.1\%$.
}
\end{figure}
maxima an minima for the case shown in \fig{figures/skposc}.
It is obvious that a fixed period is maintained over many
cycles.  This period is determined by a least squares fit
to the minimum and maximum positions.  The zero crossings
straddle the fitted line because the peak-to-peak envelope center
is still a small way below its final value.
The decay rate was determined as shown in \fig{figures/decayfit},
\def\figpath{figures/decayfit}
\begin{figure}[htb]
\centerline{\includegraphics[width=\figsize\textwidth]{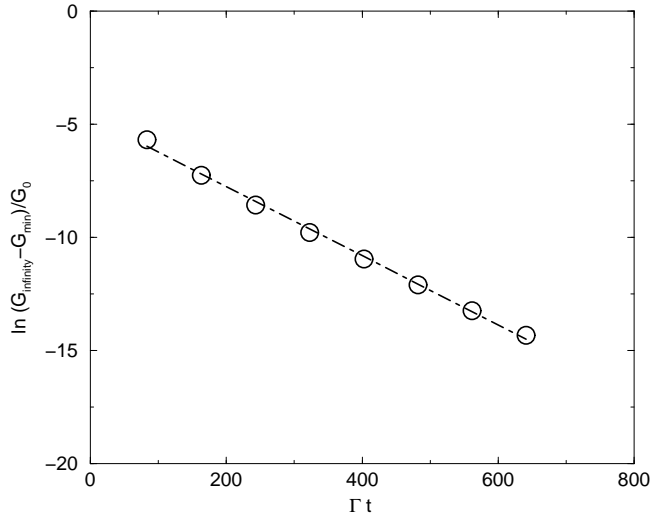}}
\caption{\label{\figpath}
Determination of the decay rate of the SKP oscillations from
the positions of the minima in Fig.~\ref{figures/skposc}.
The slope of the curve gives a rate of $0.015\Gamma$ for
$V=0.08\Gamma$ and $T=0.0015$.  These rates are universal
functions of $V/T_{\rm K}$ and $T/T_{\rm K}$, and hence
depend on $\Gamma$, $E$, and $D$ only through
\eq{tkondo}.  In the case of this figure, $T_{\rm K}=0.0022\Gamma$.
}
\end{figure}
by a least square fit to the logarithm of the difference
between the dc conductance and its value at the minimum points
vs.\ the time corresponding to these points. If one looks
closely, there appears to be a small curvature at the
beginning.  This is probably real and not noise, as
according to \fig{figures/spec2} the split Kondo peaks in the
instantaneous spectral functions are still narrowing slightly
in this region. Aside from rejecting early points that
were obviously  out of line, we did not attempt to
account for this in the analysis. The small misalignment
of the final point, on the other hand is certainly due to noise,
an ubiquitous feature for extremely large times, which
deterred us from trying to distinguish between the
'initial' and 'final' slopes.


\begin{thebibliography}{10}
\bibitem{inflection}
One can show that for large $\ln (V/T_\text{K})$,
the width of each of the split Kondo peaks in the steady-state
spectral function, as measured between the inflection
points, is the $2\gamma$ defined in Ref.~\onlinecite{rosch}.
To do this one must consider the additional rounding
mentioned there, but not included in their Eq.~(3).
This width is of course much smaller than the width at
half maximum.
\end{thebibliography}
\end{document}